\documentclass[letter]{aa}

\usepackage{graphicx}
\usepackage{txfonts}
\usepackage{natbib}
\bibpunct{(}{)}{;}{a}{}{,}    

\begin{document}

\title{Discovery of a massive supercluster system at $z\sim0.47$ }

\author{H.~Lietzen \inst{1,2} \and
  E.~Tempel \inst{3} \and
  L.~J.~Liivamägi \inst{3} \and
  A.~Montero-Dorta \inst{4} \and
  M.~Einasto \inst{3}\and
  A.~Streblyanska \inst{1,2}\and
  C.~Maraston \inst{5}\and
  J.~A.~Rubiño-Martín \inst{1,2}\and
  E.~Saar \inst{3}
}

\institute{Instituto de Astrofísica de Canarias, E-38200 La Laguna, Tenerife, Spain
  \and
  Universidad de La Laguna, Dept. Astrofísica, E-38206 La Laguna, Tenerife, Spain
  \and
  Tartu Observatory, Observatooriumi 1, 61602 T\~oravere, Estonia
  \and
  Department of Physics and Astronomy, The University of Utah, 115 South 1400 East, Salt Lake City, UT 84112, USA
  \and
ICG-University of Portsmouth, Dennis Sciama Building, Burnaby Road, PO1 3FX, Portsmouth, United Kingdom
    }

\abstract
    {}
    {Superclusters are the largest relatively isolated systems in the cosmic web. Using the SDSS BOSS survey we search for the largest superclusters in the redshift range $0.43<z<0.71$.}
    {We generate a luminosity-density field smoothed over $8~h^{-1}\mathrm{Mpc}$ to detect the large-scale over-density regions.  Each individual over-density region is defined as single supercluster in the survey. We define the superclusters in the way that they are comparable with the superclusters found in the SDSS main survey.}
    {We found a system we call the BOSS Great Wall (BGW), which consists of two walls with diameters 186 and 173~$h^{-1}$Mpc, and two other major superclusters with diameters of 64 and 91~$h^{-1}$Mpc. As a whole, this system consists of 830 galaxies with the mean redshift 0.47. We estimate the total mass to be approximately $2\times10^{17}h^{-1}M_\odot$. The morphology of the superclusters in the BGW system is similar to the morphology of the superclusters in the Sloan Great Wall region.}
    {The BGW is one of the most extended and massive system of superclusters yet found in the Universe.}

    \keywords{Large-scale structure of the Universe}

    \maketitle
    \section{Introduction}

The large-scale structure of the Universe can be seen as the cosmic web of clusters and groups of galaxies connected by filaments, with under-dense voids between the over-dense regions \citep{1996Natur.380..603B}. The largest over-dense, relatively isolated systems in the cosmic web are the superclusters of galaxies \citep{deVaucouleurs1956, Joeveer1978, Zucca1993, Einasto1994}. 

    Several supercluster catalogs have been compiled recently \citep{Einasto2007, Liivamagi2012, Chow-Martinez2014, Nadathur2014}, providing material for studies on the large-scale structure. We are now extending the knowledge of superclusters to the redshifts above 0.45, using the CMASS (constant mass) sample of the Sloan Digital Sky Survey III (SDSS-III) \citep{Eisenstein2011,Maraston2013,Reid2016}. Only a few relatively small superclusters have been found at high redshifts before \citep{Tanaka2007, Schirmer2011, Pompei2015}. 

We have found an unusually extended overdensity within the SDSS/CMASS volume, at redshift $z\sim0.47$. This structure resembles the Sloan Great Wall, which consists of several superclusters and is the richest and largest system found in the nearby universe \citep{Vogeley2004,Einasto2011}. Another comparison point is the local Laniakea supercluster with 160~Mpc diameter \citep{Tully2014, 2014Natur.513...41T}.

    Throughout this paper, we assume a $\Lambda$CDM cosmology with total matter density $\Omega_{\mathrm{m}}=0.27$, dark energy density $\Omega_\Lambda=0.73$. We express the Hubble constant as $H_0=100$\,$h$\,km\,s$^{-1}$Mpc$^{-1}$\citep{Komatsu2011}.
    
\section{Data}

We use data from the twelfth data release (DR12) of the SDSS \citep{Alam2015,York2000} Baryon Oscillation Spectroscopic Survey \citep[BOSS;][]{Eisenstein2011, Bolton2012, Dawson2013}. We use the CMASS (constant mass) sample, which targets galaxies in the redshift range $0.43<z<0.7$. The BOSS data is obtained using a multi-object spectrograph \citep{Smee2013} on the 2.5-m telescope \citep{Gunn2006} located at Apache Point Observatory in New Mexico. The SDSS imaging was done with a drift-scanning mosaic CCD camera \citep{Gunn1998} in five color-bands, $u,g,r,i,z$ \citep{Fukugita1996}, and it was published in Data Release~8 \citep[DR8;][]{Aihara2011}. BOSS obtains spectra with resolution of 1500 to 2600 in the wavelength range 3600 to 10\,000~\AA\ \citep{Smee2013}.

The CMASS sample selects massive and luminous galaxies at redshift above $z\sim 0.4$, whose stellar mass remains approximately constant up to $z\sim 0.6$. In principle, the selection criteria allow the detection of galaxies with arbitrary colors, as the CMASS cut does not pose any limit to the observed-frame $g-r$. In practice, the CMASS cut selects the massive end of the red sequence, as these are the most abundant galaxies at the high mass end ($M>10^{11}~M_{\odot}$) and they do not evolve over the CMASS redshift range.
The CMASS sample was selected to isolate the massive end of the red sequence at $z\sim0.5$.
First, a pre-selection was performed to ensure that the targets pass a set of quality criteria described in \citet{Dawson2013}. The final selection was made based on the observed colors and magnitudes \citep{Cannon2006, Abazajian2009, Anderson2012}.

For a comparison to the low-redshift universe, we used the SDSS DR7 main sample superclusters from \citet{Liivamagi2012}. This sample consists of 583\,362 galaxies in a volume of $1.32\times 10^8$\,$h^{-3}$Mpc$^{3}$, with mean number density $4.4\times 10^{-3}$\,$h^3$Mpc$^{-3}$. The comparison sample includes nearly 1000 superclusters.

\section{Methods}

We constructed a luminosity-density field to distinguish the superclusters from the lower-density regions following the same procedure that was used in \citet{Liivamagi2012}. We weighted the luminosities of galaxies to set the mean density the same through the whole distance range, and then calculated the density field in a 3\,$h^{-1}$Mpc grid with an 8\,$h^{-1}$Mpc smoothing scale.

\begin{figure}[t]
\centering
\resizebox{\hsize}{!}{\includegraphics{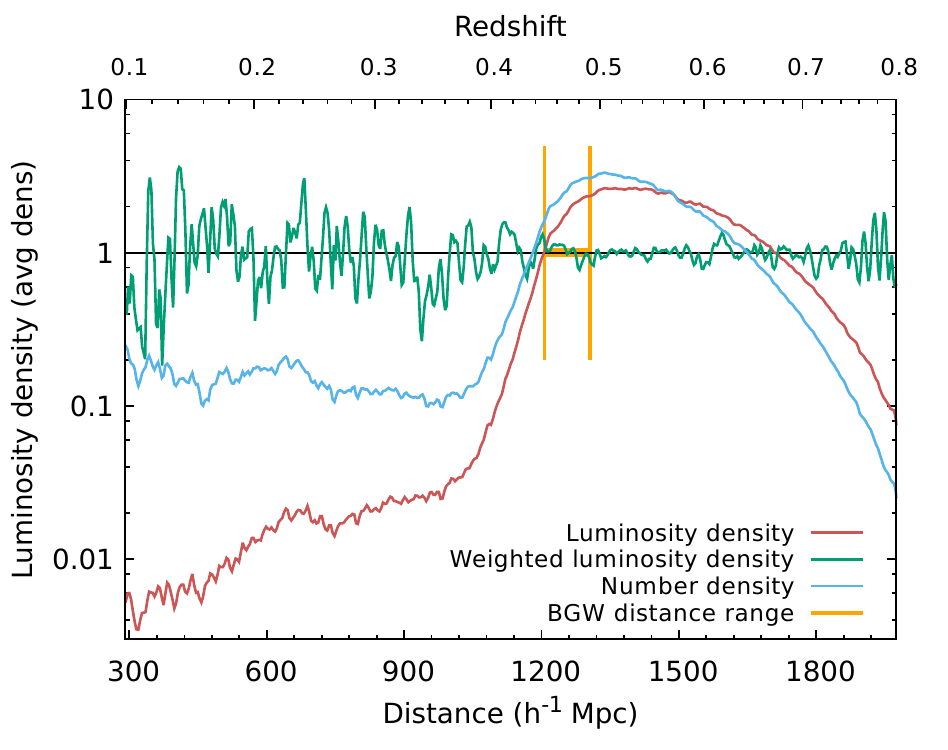}}
\caption{Number density (blue), luminosity density (red), and weighted luminosity density (green) of the CMASS galaxies as a function of distance. The densities are given in units of mean density. The orange region shows the discovered BGW region.}
\label{densdist}
\end{figure}

The number density and the luminosity density of the galaxies varies with distance as shown in Fig.~\ref{densdist}. The weighting suppresses this variation, making the mean luminosity density remain constant with small random variation throughout the distance range from 1200 to 1800\,$h^{-1}$Mpc. We also trimmed the edges of the sample in the survey coordinates with the limits $-50.0\degr<\lambda<51.5\degr$, $-34.5\degr<\eta<36.25\degr$, where $\lambda$ and $\eta$ are the SDSS
survey coordinates. In the SDSS, the survey coordinates form a spherical coordinate system, where $(\eta,\lambda) = (0,90.)$ corresponds to $(R.A.,Dec.) = (275.,0.)$, $(\eta,\lambda) = (57.5,0.)$ corresponds to $(R.A.,Dec.) = (0.,90.)$, and at  $(\eta,\lambda) = (0.,0.)$,  $(R.A.,Dec.) = (185.,32.5)$. Volume of the luminosity-density field with these limits is $2.62\times 10^9$\,$h^{-3}$Mpc$^{3}$, and it contains 480801 galaxies, making the mean number density of galaxies $1.8\times 10^{-4}$\,$h^3$Mpc$^{-3}$.

\begin{figure}[t]
                \centering
\resizebox{\hsize}{!}{\includegraphics{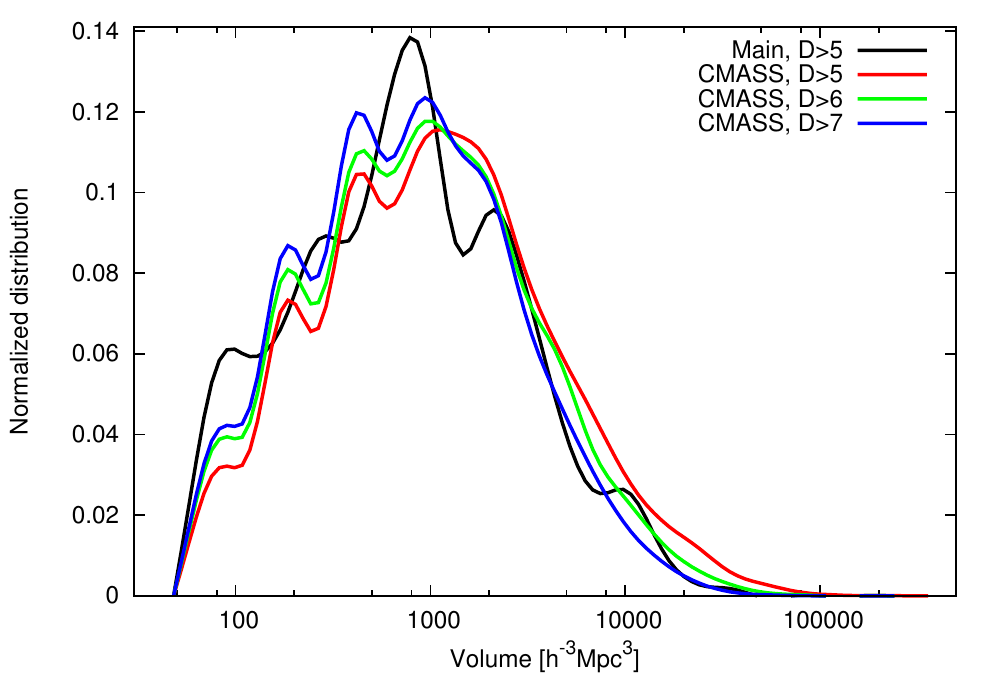}}
        \caption{Distribution of supercluster volumes. For comparison, the distribution of the SDSS DR7 main sample is shown with the black curve. The CMASS supercluster volume distribution is shown for three density thresholds (in units of mean density): $D>5$ (red), $D>6$ (green), and $D>7$ (blue).}
        \label{voldist}
        \end{figure}

      In previous studies on the SDSS main sample, the threshold density of 5.0 times the mean density has often been used as the limit for superclusters \citep[e.g.][]{2011A&A...529A..53T, Lietzen2012, Einasto2014}. We use the main sample superclusters from \citet{Liivamagi2012} as a comparison to the superclusters in the CMASS sample. Figure~\ref{voldist} shows the volume distribution for superclusters found with density thresholds of five, six, and seven mean densities compared to the SDSS main sample in DR7 with density threshold of five times the mean density. The volume was calculated as the number of grid cells multiplied by the volume of one cell. Since the grid size in the CMASS density field is 3\,$h^{-1}$Mpc, the volume of a grid cell is 27\,$(h^{-1}\mathrm{Mpc})^3$.\footnote{The grid size of the main SDSS sample field is 1\,$h^{-1}$Mpc. Since we smooth the density field in both, SDSS and CMASS, samples with an 8\,$h^{-1}$Mpc kernel, the grid size of 3\,$h^{-1}$Mpc is equally comparable with grid size of 1\,$h^{-1}$Mpc.}  The distributions of supercluster volumes with these three thresholds are close to that of the main sample superclusters, suggesting that the density limit to determine superclusters should be around these density levels.

    \section{Results}

The largest structure that we found with the density threshold $D>5$ mean densities, has a diameter of 271\,$h^{-1}$Mpc and a volume of $2.4\times 10^5 (h^{-1}\mathrm{Mpc})^3$, and contains 830 galaxies. The richness difference to the other structures is significant, as the second richest one contains only 390 galaxies. The structure of the largest system is more complex than the structure of the SDSS main sample superclusters, containing several separate cores. This suggests that structures found with this density threshold are not individual superclusters, but systems of several separate superclusters. This system can be seen comparable to the Sloan Great Wall, which is also a system of several superclusters \citep{Einasto2011}.

\begin{figure}
\centering
\resizebox{\hsize}{!}{\includegraphics{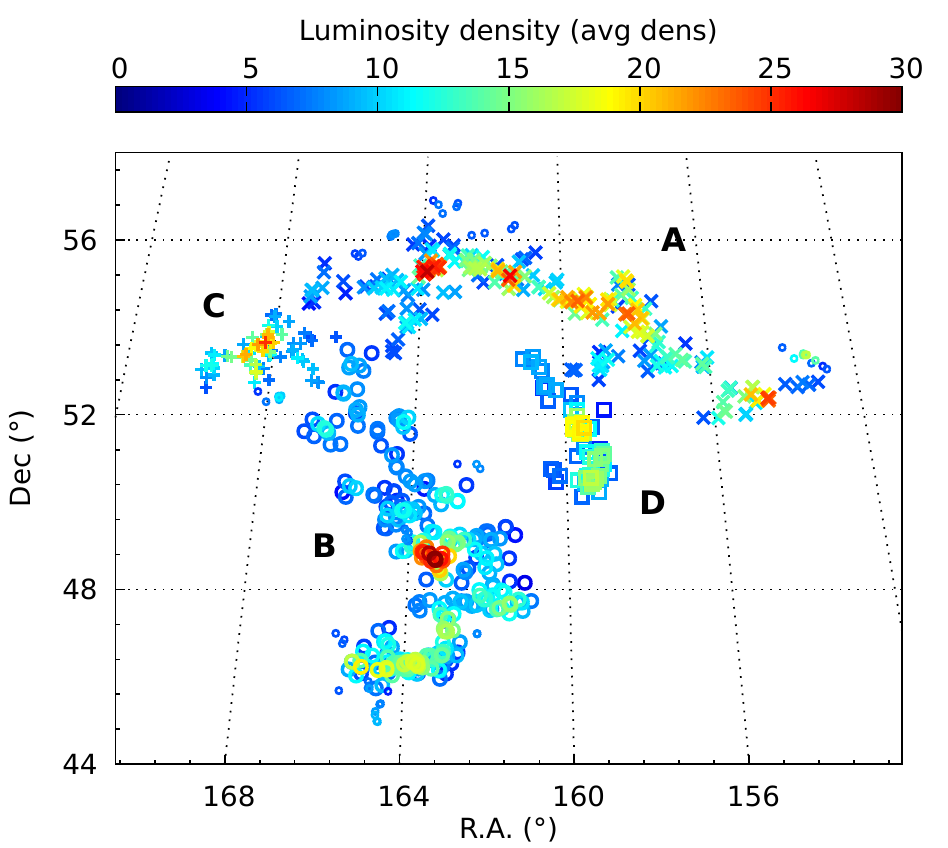}}
\caption{Galaxies in the BOSS Great Wall (BGW) in sky coordinates. The color scale shows the local environmental density in terms of mean densities for each galaxy. The different symbols refer to galaxies in the four largest superclusters in the system determined with the density threshold $D>6$.}
\label{bigsky}
\end{figure}

When we increase the density threshold, the unusually large overdensity found with $D>5$ level breaks in several parts. With the density level $D>6$ mean densities, the individual superclusters can be distinguished from each others. The structure is shown in sky coordinates in Fig.~\ref{bigsky}, with the galaxies in the four largest superclusters shown with filled symbols, and other galaxies belonging to the $D>5$ overdensity shown with crosses. The most prominent feature of the structure are two walls with diameters 186\,$h^{-1}$Mpc (supercluster A in Fig.~\ref{bigsky}) and 173\,$h^{-1}$Mpc (supercluster B in Fig.~\ref{bigsky}). Again, these are the two largest superclusters in the whole sample by diameter. In addition, there are two moderately large superclusters (marked as C and D in Fig.~\ref{bigsky}) with diameters 64 and 91\,$h^{-1}$Mpc and a few small superclusters with less than 15 galaxies within this structure.

We analysed the shapes of the two richest superclusters
A and B using the shape parameter K which is defined
with Minkowski functionals \citep[see][for details]{2011A&A...532A...5E}.
This is the ratio
of two parameters, planarity and filamentarity, which both
may change from 0 to 1 (from a sphere to a plane or to a
line). So, the smaller the parameter, the more elongated a
system is. The superclusters A and B are very elongated, with
the shape parameter values 0:17 and 0:19. For a comparison, the richest
supercluster in the
Sloan Great Wall has the shape parameter K = 0:27, and is
therefore less elongated than the two richest superclusters in
the BGW.

\begin{figure}
\centering
\resizebox{\hsize}{!}{\includegraphics{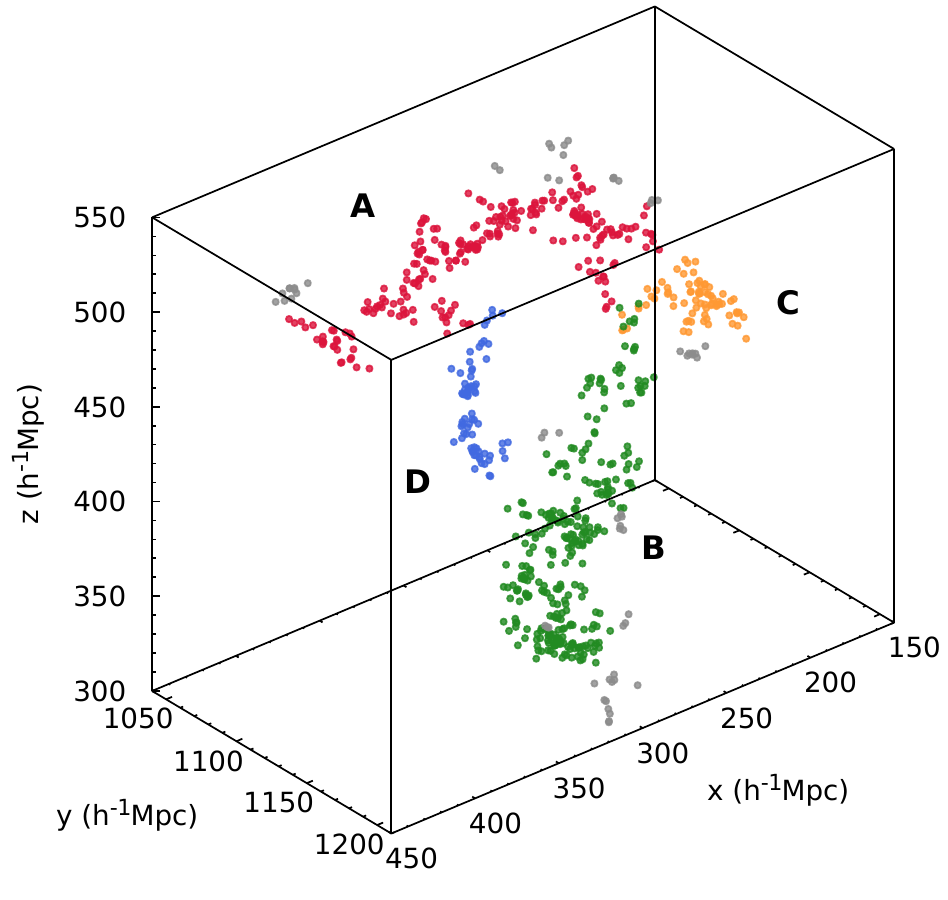}}
\caption{Galaxies in the BGW superclusters in Cartesian coordinates. Different colors show the individual superclusters in the BGW system.}
\label{big3d}
\end{figure}

We will hereafter call this system the BOSS Great Wall (BGW). In Fig~\ref{big3d} the supercluster galaxies are shown in the three-dimensional Cartesian coordinates which were defined as
      \begin{align}
      x&=-d\sin{\lambda}\\
      y&=d\cos{\lambda}\cos{\eta}\\
      z&=d\cos{\lambda}\sin{\eta},
      \end{align}
      where $d$ is the distance of the galaxy and $\eta$ and $\lambda$ are the SDSS angular coordinates. The basic information about the two walls and the two companion superclusters is given in Table~\ref{BigScls}. At \url{http://www.aai.ee/~maret/BOSSWalls.html}
we present an interactive 3D model showing the distribution of
galaxies in the BOSS Great Wall.

\begin{table*}
\centering
\begin{tabular}{c c c c c c c c}
\hline\hline
Supercluster & Richness & Diameter  & Volume & Average density & Maximum density & $\log(M_1)$ & $\log(M_2)$\\
 & $N_{\mathrm{gal}}$ & $h^{-1}$Mpc & ($h^{-1}$Mpc)$^3$ & $\langle \rho_\mathrm{L} \rangle$ & $\langle \rho_\mathrm{L} \rangle$ & $\log(h^{-1}M_\odot)$ & $\log(h^{-1}M_\odot)$\\
\hline       
A & 255 & 186.1 & 67500 & 9.1 & 27.9 & 16.8 & 16.9\\
B & 303 & 172.9 & 70848 & 9.3 & 29.6 & 16.6 & 16.9\\
C & 73 & 63.8 & 19008 & 10.2 & 24.5 & 16.3 & 16.4\\
D & 71 & 90.6 & 13635 & 9.3 & 21.2 & 16.0 & 16.2\\
\hline
Total & 830 & 271.1 & 241029 & 8.6 & 29.6 & 17.2 & 17.4\\
\hline
\end{tabular}
\caption{Basic information on the main superclusters in the BOSS Great Wall supercluster system and the whole system. The line ``Total'' refers to the over-dense region found with the threshold of $D>5$ mean densities. The individual superclusters in the BGW are found with the threshold $D>6$. The two mass estimates are the mass derived from the stellar masses ($\log(M_1)$) and the mass derived from the critical density ($\log(M_2)$).}
\label{BigScls}
\end{table*}

\subsection{Total mass of the BGW superclusters}

We used stellar masses to estimate a minimum mass for the BGW and its member superclusters.
BOSS stellar masses are obtained from the Portsmouth galaxy product 
\citep{Maraston2013}, which is based on the stellar population 
models of \cite{Maraston2005} and \cite{Maraston2009}. The Portsmouth 
product uses an adaptation of the publicly-available Hyper-Z code 
\citep{Bolzonella2000} to perform a best-fit to the observed {\it{ugriz}} 
magnitudes of BOSS galaxies, with the spectroscopic redshift determined by the BOSS pipeline. 
The stellar masses that we use in this work were computed assuming 
a Kroupa initial mass function. 

Stellar masses of galaxies in superclusters are typically higher than at low densities \citep{Einasto2014}. Similarly, in the BGW system, the stellar masses of galaxies in the two walls are higher than the stellar masses in a control sample of galaxies with the local density level $D<2$ times the mean density, with median values $9.1 \times 10^{10}$ for supercluster A,  $8.3 \times 10^{10}$ for supercluster B, and $6.9 \times 10^{10}$ for the low-density galaxies. According to the Kolmogorov-Smirnov (KS) test, both walls have a stellar mass distribution that differs from the low-density galaxies to very high level of significance: $p$-values are less than $10^{-10}$. For the difference between the two walls the KS test gives the $p$-value of 0.0329, which means that the stellar mass distributions are  different with the significance level of 5\,\%.

We estimate a lower limit for the mass of the walls using the stellar mass $M_*$ to halo mass $M_{\mathrm{halo}}$ relation as given by \citet{Moster2010}, assuming that all galaxies in the CMASS sample are central galaxies of a halo.
According to \citet{Maraston2013}, the sample is roughly complete at stellar masses above $10^{11.3}h^{-1}M_\odot$ in our redshift range \citep[see also][]{Leauthaud2015}. The sample therefore misses the halos hosting only low-mass galaxies due to the data selection. We correct for this incompleteness by scaling the SDSS main sample of galaxies as a comparison. For this we used the magnitude-limited galaxy and friend-of-friend (FoF) group catalog by \citet{Tempel2014} in the distance bin from 180 to 270\,$h^{-1}$Mpc. In this sample, the ratio between stellar mass in BOSS-like sample of galaxies with $\log(M_*/h^{-1}M_\odot)>11.3$ and the most luminous galaxies of the FoF groups in superclusters is 0.082. Correcting the mass in the BGW superclusters with this ratio, we get masses $5.9\times10^{16}h^{-1}M_\odot$ for supercluster A, $4.4\times10^{16}h^{-1}M_\odot$ for supercluster B, and $1.6\times10^{17}h^{-1}M_\odot$ for the whole system.

Another way to estimate the total mass of the superclusters is to use the critical density of the Universe $\rho_c$. We assumed that mass density correlates with luminosity density, and calculated the mass of each grid cell belonging to the superclusters as $M=\rho_c D V$, where $D$ is the luminosity density in the units of mean densities and $V=27~h^{-3}$Mpc$^3$ is the volume of the cell. With this method we get masses $8.2\times10^{16}h^{-1}M_\odot$ for supercluster A and $7.7\times10^{16}h^{-1}M_\odot$ for supercluster B. The mass of the BGW in total is $2.4\times10^{17}h^{-1}M_\odot$. The mass estimates for all BGW superclusters are shown in Table~\ref{BigScls}.

\section{Discussion and conclusions}

We found two walls of galaxies at redshift $0.45<z<0.5$ that are larger in volume and diameter than any previously known superclusters. Together they form the system of the BOSS Great Wall, which is more extended than any other known structure. The closest comparison to this system is the Sloan Great Wall, which is also a system of several large superclusters, and complexes of superclusters connected with the Sloan Great Wall \citep{2011A&A...532A...5E}. However, the volume of the main supercluster of the SGW is
$2.5\times 10^4$\,($h^{-1}$Mpc)$^3$, which is smaller than the volumes of either of the walls in the BGW \citep{Einasto2011}. The BGW system as a whole covers a volume of $2.4\times 10^5$\,($h^{-1}$Mpc)$^3$ of which $1.7\times 10^5$\,($h^{-1}$Mpc)$^3$ consists of the four largest superclusters.
With the diameter of 271\,$h^{-1}$Mpc, the BGW is considerably more extended than the SGW which has a diameter of approximately 160\,$h^{-1}$Mpc \citep{Sheth2011} or the local Laniakea supercluster, whose full basin of attraction has the diameter of 160\,$h^{-1}$Mpc \citep{Tully2014}.

Stellar masses have not previously been used to estimate the total masses of superclusters. Our mass estimates from stellar masses agree well with our estimate from critical density. Our analysis suggests that both of the two walls in the BGW are comparable to the entire SGW, whose mass is estimated to be within 20\,\% of $1.2\times10^{17}h^{-1}M_\odot$ \citep{Sheth2011}.  We can therefore conclude that the total mass of the BGW system is approximately $2\times 10^{17}h^{-1}M_\odot$, making the BGW the most massive system of superclusters found in the Universe.

    \begin{acknowledgements}

The authors thank Brent Tully, for his valuable suggestions for improving the article as a referee.

HL, AS, and JAR-M  acknowledge financial support from the Spanish Ministry
of Economy and Competitiveness (MINECO) under the 2011 Severo
Ochoa Program MINECO SEV-2011-0187.

ET, LJL, ME, and ES were supported by the ETAG projects IUT26-2, IUT40-2, and TK133.

AMD acknowledges support from the U.S. Department of Energy, Office of Science, Office of High Energy Physics, under Award Number DE-SC0010331. AMD also thanks the Center for High Performance Computing at the University of Utah for its support and resources.

Funding for SDSS-III has been provided by the Alfred P. Sloan Foundation, the Participating Institutions, the National Science Foundation, and the U.S. Department of Energy Office of Science. The SDSS-III web site is http://www.sdss3.org/.

SDSS-III is managed by the Astrophysical Research Consortium for the Participating Institutions of the SDSS-III Collaboration including the University of Arizona, the Brazilian Participation Group, Brookhaven National Laboratory, Carnegie Mellon University, University of Florida, the French Participation Group, the German Participation Group, Harvard University, the Instituto de Astrofisica de Canarias, the Michigan State/Notre Dame/JINA Participation Group, Johns Hopkins University, Lawrence Berkeley National Laboratory, Max Planck Institute for Astrophysics, Max Planck Institute for Extraterrestrial Physics, New Mexico State University, New York University, Ohio State University, Pennsylvania State University, University of Portsmouth, Princeton University, the Spanish Participation Group, University of Tokyo, University of Utah, Vanderbilt University, University of Virginia, University of Washington, and Yale University. 

    \end{acknowledgements}


\end{document}